\documentclass[aps,prb,superscriptaddress,twocolumn,longbibliography]{revtex4-2}

\usepackage[T1]{fontenc}
\usepackage{lmodern}
\usepackage[svgnames]{xcolor}
\usepackage{lipsum}

\definecolor{Dark}{gray}{0.2}
\definecolor{MedDark}{gray}{0.4}
\definecolor{Medium}{gray}{0.6}
\definecolor{Light}{gray}{0.8}
\definecolor{darkred}{rgb}{0.55, 0.0, 0.0}
\definecolor{darkslateblue}{rgb}{0.28, 0.24, 0.55}
\definecolor{royalblue(web)}{rgb}{0.25, 0.41, 0.88}
\usepackage{graphicx,float}
\usepackage{amsmath}
\DeclareMathOperator{\Tr}{Tr}
\usepackage{amssymb}
\usepackage[colorlinks]{hyperref}
\hypersetup{
    citecolor=darkred,
    linkcolor=blue,   
    urlcolor=Magenta}

\usepackage{bbm}
\usepackage[english]{babel}

\def\A{\mathrm{A}}
\def\B{\mathrm{B}}
\def\u{\uparrow}
\def\d{\downarrow}

\def\hamil{{\mathcal{H}}}
\def\lbk#1{\left(#1\right)}

\def\meanvl#1{\langle #1 \rangle}

\def\be{\begin{equation}}
\def\ee{\end{equation}}
\def\beq{\begin{equation}}
\def\eeq{\end{equation}}
\def\bea{\begin{eqnarray}}
\def\eea{\end{eqnarray}}
\def\nbea{\begin{eqnarray*}}
\def\neea{\nonumber\end{eqnarray*}}
\def\bmat#1{\left(\begin{array}{#1}}
\def\emat{\end{array}\right)}
\def\bcase#1{\left\{\begin{array}{#1}}
\def\ecase{\end{array}\right.}
\def\bmini#1{\begin{minipage}{#1\textwidth}}
\def\emini{\end{minipage}}

\usepackage{wrapfig}
\graphicspath{{figures/}}
\usepackage{enumerate}
\newcommand{\cpic}[2]{
\includegraphics[scale=#1]{#2}
}

\newcommand{\cpicn}[3]
{
\begin{figure}
\begin{center}
\cpic{#1}{#2}
\caption{#3\label{#2}}
\end{center}
\end{figure}
}

\begin{document}
\title{Creating long-range entangled Majorana pairs: from spin-1/2 twisted Kitaev to generalized XY chains}
\date{\today} 
\author{Haoting Xu}
\affiliation{Department of Physics, University of Toronto, 60 St. George St., Toronto, Ontario, Canada M5S 1A7}
\author{Hae-Young Kee}
\email[]{hykee@physics.utoronto.ca}
\affiliation{Department of Physics, University of Toronto, 60 St. George St., Toronto, Ontario, Canada M5S 1A7}
\affiliation{Canadian Institute for Advanced Research, CIFAR Program in Quantum Materials, Toronto, Ontario, Canada M5G 1M1 }

\begin{abstract}
Entangling a pair of far-distant qubits in many-body systems has been a challenging task in quantum computing. A robust entanglement was predicted in the rainbow states and generating nonlocal Bell pairs protected by a mirror symmetry was recently proposed. 
We investigate a way to create entangled Majorana fermions in the spin-1/2 Kitaev chain with open boundary conditions. The spin-1/2 Kitaev chain, a one-dimensional version of the honeycomb lattice with bond-dependent Ising interactions, has a macroscopic degeneracy related to the zero modes containing non-local spin strings.
We show that applying a pair pulse sequence on the central unit cell of the chain promotes long distance spin correlations and maximal bipartite entanglement entropy. We extend this method to the generalized bond-dependent spin-1/2 chain by introducing another set of Majorana fermions and
make a comparison to the entangled Bell pairs.
The time to reach maximal bipartite entanglement entropy is shorter in the Kitaev chain as the zero modes do not participate in the entangled pairs. 
An application of our results to a recently proposed twisted Kitaev chain, CoNb$_2$O$_6$, is presented and future directions are also discussed.
\end{abstract}

\maketitle

\section{Introduction\label{intro}}
The notion of entanglement, inherent non-locality of quantum information, has become the heart of modern quantum technology, as generating long-range entangled states is a fundamental ingredient of quantum computing and quantum networks~\citep{divincenzo2000physical,preskill1998quantum,steane1998quantum,wehner2018quantum,gottesman1999demonstrating,gyongyosi2019survey,ladd2010quantum,de2021materials,valiev2005quantum,aharonov1999quantum,RevModPhys.68.733,vedral1998basics,briegel2009measurement,buluta2011natural,raussendorf2012quantum,bennett2000quantum,kilin1999quantum,kimble2008quantum,IEEE1,IEEE2}.
One of the most well-known entangled states 
is the Bell pair state, which is a non-direct product state of two qubits~\citep{RevModPhys.81.865}. However, decoherence of entangled pairs of two distant qubits 
can be caused by interactions in many body systems.~\citep{chirolli2008decoherence,kendon2002typical,brandt1999qubit,schlosshauer2005decoherence} 
In order to overcome the decoherence problem, 
the rainbow state ~\citep{Ramirez_2014,
PhysRevA.77.020303,PhysRevA.78.012330,PhysRevA.90.042304,pitsios2017photonic,PhysRevB.105.L140301,PhysRevLett.125.240404,dutta2022generating,vitagliano2010volume,PhysRevResearch.3.L012016,persistent2001Briegel,ramirez2015entanglement,rodriguez2017more}, a direct product of two-site entangled states, has been suggested. An $N$-site rainbow state requires measurement of $N/2$ sites to disentangle it, which implies the state has a high persistency, defined by the minimal number of local measurements to completely disentangle the state. ~\citep{persistent2001Briegel} 
Although the decoherence with the environment may still be an issue in the rainbow state unlike topological states ~\citep{topological_state_haldane}, 
the rainbow state following the volume law~\citep{Ramirez_2014,PhysRevB.105.L140301} contains end-to-end deterministic quantum entanglement, which makes it suitable for entanglement distribution networks~\citep{wehner2018quantum}. There have been intensive studies to generate the rainbow state. 
Methods of generating the rainbow state include optimizing the coupling strength of the system~\citep{persistent2001Briegel,
vitagliano2010volume,Ramirez_2014,PhysRevA.77.020303,PhysRevA.78.012330}, quenching the selective interactions~\citep{PhysRevA.90.042304,pitsios2017photonic}, and using the dissipation between the system and a reservoir~\citep{PhysRevB.105.L140301,PhysRevLett.125.240404,PhysRevResearch.3.L012016}.

Recently, ~\citet{PhysRevLett.125.240404} and ~\citet{dutta2022generating} proposed a variant of the rainbow state
which is protected by a mirror symmetry of the system. 
Their results have not only reflected the significant role of symmetry in quantum computing problems~\citep{gross1996role}, but also provided more insights and clues towards how to manipulate target systems. ~\citet{dutta2022generating} further
proposed a protocol based on the symmetry of generating the rainbow-like state by applying a sequence of simultaneous $\pi$-pulses on half of the spin-1/2 XX chain, which can be done in today's cold-atom experiments. 
In this paper, we investigate the possibility of creating rainbow-like states in solid-state materials. We propose that CoNb$_2$O$_6$ is a promising candidate for realizing the rainbow state. Recent studies suggest that CoNb$_2$O$_6$, previously thought to be an Ising chain, is better described by a twisted Kitaev chain model ~\citep{morris2021duality}. While the Kitaev spin-1/2 chain contains alternating nearest neighbor Ising interactions between the bonds (such as $S^x S^x $ and $S^y S^y$), the twisted Kitaev model modifies this by replacing the ${\hat x}$ and ${\hat y}$ spin directions with the $\hat{n}_1$ and $\hat{n}_2$ directions, which are not necessarily perpendicular to each other. 

To understand the dynamics of the twisted Kitaev chain, we begin by examining the spin-1/2 Kitaev spin chain, which is a 1D limit of the Kitaev honeycomb model~\citep{kitaev2006anyons,QPT_of_Kitaev}, also known as the quantum compass chain ~\citep{You2008compass,Brzezicki2007compass,sun2009compass,You2014compass,Eriksson2009compass,laurell2022spin}.
The quantum compass chain can be expressed in free fermion form using the Jordan-Wigner transformation. Nevertheless, due to the non-local $N/2$ SU(2) symmetries of the model where $N$ is a number of sites\citep{Sen2010PRB}, there are $2^{N/2}$ macroscopic degeneracy resulting in novel physics such as Majorana zero modes ~\citep{You2008compass,Brzezicki2007compass,sun2009compass,You2014compass,Eriksson2009compass,laurell2022spin,agrapidis2018ordered}, hidden string order parameter ~\citep{QPT_of_Kitaev}, and divergence of the structure factor ~\citep{laurell2022spin}. Thus a protocol to create the rainbow state in this model may differ from the previously studied spin-1/2 XX chain ~\citep{dutta2022generating,PhysRevB.105.L140301,PhysRevLett.125.240404,vitagliano2010volume,PhysRevResearch.3.L012016}.
We will show below that entangled Majorana fermions between the mirror symmetric sites can be created by a sequence of the pair pulses.
%
We also study the time to reach the far-distant (size of the system) entangled pairs in the Kitaev chain and compare that to the generalized XY spin-1/2 chain. 
We find the Kitaev chain takes a short time to reach the maximal bipartite entanglement than that of the generalized XY-chain, most likely due to the fact that the zero modes in the Kitaev chain is inactive. We show how to transform the entangled Majorana fermions to the entangled Bell pair in spin language. We apply our results to  CoNb$_2$O$_6$~\cite{morris2021duality,laurell2022spin}, and show that qualitatively similar results hold for the twisted system.

The rest of this paper is organized as follows.
In Sec.~\ref{sec:Kitaev} we introduce the Kitaev model and describe it in terms of the Majorana fermions. With a quick review of non-locality of the model, we identify a conserved quantity, and develop a process of generating the long-range pairs of Majorana fermions. Numerical results of the time-dependence of spin correlation and bipartite entanglement entropy are also presented. In Sec.~\ref{sec:XY}, we extend our method to the generalized XY spin chain, where another set of Majorana fermions is needed to describe the full Hamiltonian. In Sec.~\ref{sec:Bell} we show the connection between the entangled Majorana pairs and the Bell pairs. In Sec.~\ref{sec:flip} we discuss an alternative process of generating a fraction of the maximum entanglement by applying the $\pi$-pulse, i.e., flipping  spins of half of the chain. In Sec.~\ref{sec:twisted} we apply our method to the twisted Kitaev chain and show that maximal entanglement entropy can be reached despite its deviation from the ideal Kitaev chain. In Sec.~\ref{sec:stability} we present the effects of various perturbations on the entanglement entropy of the rainbow state. In Sec.~\ref{sec:Diss} we summarize our results and discuss the limitation and extension of our theory to related systems.

\section{Kitaev chain\label{sec:Kitaev}}
\begin{figure}
	\includegraphics[width = 0.5\textwidth]{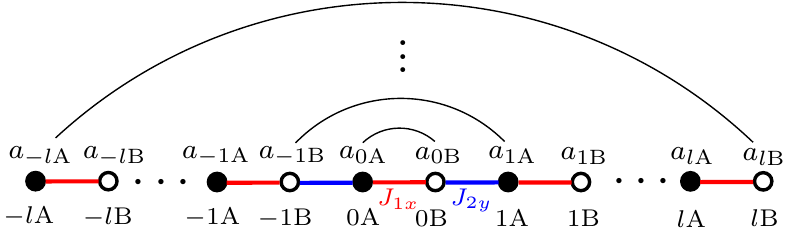}
	\caption{The description of Kitaev chain in terms of Majorana fermion $a_{i\mu}$. The red and blue bonds represent $J_{1x}$ and $J_{2y}$ interactions, respectively. The Hamiltonian of Kitaev chain can be expressed with nearest neighbor hopping of $a_{i\mu}$, see Eq.~\ref{eq:hamil_kitaev_marjoana}. The arcs show the entangled Majorana pairs in the rainbow-like state.}
	\label{fig:definition_kitaev}
\end{figure}

We consider the spin-$\frac12$ bond-dependent Kitaev chain
with an odd integer $(2l +1)$ number of unit cells, i.e., $N = 2 (2 l +1)$ total sites, as shown in Fig.~\ref{fig:definition_kitaev}.
The Hamiltonian is given by
\beq 
\label{eq:hamil_kitaev}
\hamil = \lbk{\frac{-J_{1x}}{4}} \sum_{j=-l}^l \sigma_{j\mathrm{A}}^x \sigma_{j\mathrm{B}}^x + \lbk{\frac{-J_{2y}}{4}} \sum_{j =-l}^{l-1} \sigma_{j\mathrm{B}}^y \sigma_{j+1, \mathrm{A}}^y 
\eeq 
where $\sigma_{j\mu}$ is the Pauli matrix on the $\mu$ sublattice $(\mu,\nu = \mathrm{A,B})$ of the $j$-th unit cell.
%
To represent $\hamil$ in terms of Majorana fermions, we introduce a complex fermion, $f_{j\mu}$, defined by $f_{j\mathrm{A}} = \sigma^-_{j\mathrm{A}}\prod_{n<j}\sigma^z_{n\mathrm{A}}\sigma^z_{n\mathrm{B}}$, $f_{j\mathrm{B}} = \sigma^-_{j\mathrm{B}}\sigma^z_{j\mathrm{A}}\prod_{n<j}\sigma^z_{n\mathrm{A}}\sigma^z_{n\mathrm{B}}$, with $\sigma^\pm_{j\mu} = (\sigma_{j\mu}^x\pm i\sigma_{j\mu}^y)/2$.  
For each pair of fermionic operators, we define two Majorana fermion operators as follows,
\begin{eqnarray}
a_{j\mathrm{A}} &=& i(f_{j\mathrm{A}}^\dagger - f_{j\mathrm{A}}), \;\;
a_{j\mathrm{B}} =f_{j\mathrm{B}}+f_{j\mathrm{B}}^\dagger,\nonumber\\
b_{j\A} &=& f_{j\A}+f_{j\A}^\dagger, \;\; 
b_{j\B} = i(f_{j\B}^\dagger - f_{j\B}).
\end{eqnarray}
These Majorana fermions 
 satisfy $\{a_{j\mu},a_{n\nu}\} = 2\delta_{jn}\delta_{\mu\nu}\mathbbm{1}$, $\{b_{j\mu},b_{n\nu}\} = 2\delta_{jn}\delta_{\mu\nu}\mathbbm{1}$ and all other anti-commutation relations are zero. This Majorana description of the Kitaev chain is plotted in Fig.~\ref{fig:definition_kitaev}. 
Then, $\hamil$  can be represented by nearest-neighbor hopping of Majorana fermions:
\beq 
\label{eq:hamil_kitaev_marjoana}
\hamil = \frac{iJ_{1x}}{4} \sum_{j = -l}^{l} a_{j\mathrm{B}} a_{j\mathrm{A}}  + \frac{iJ_{2y}}{4} \sum_{j = -l}^{l-1} a_{j\mathrm{B}} a_{j+1,\mathrm{A}} .
\eeq 

To generate  entangled Majorana pairs, we first identify a conserved quantity associated with Majorana pairs between the mirror $(j\A, -j\B)$ or $(j\B, -j\A)$ sites. It can be proven using the dynamic equations in Appendix ~\ref{appendix:conserve} that  $[\mathcal{C}_\alpha,\hamil]=0$, where the $\mathcal{C}_\alpha$ is defined by
\beq
\label{eq:C_alpha}
\mathcal{C}_\alpha = \underbrace{\frac{i}{2} a_{0\mathrm{B}} a_{0\mathrm{A}} }_{\mathcal{C}_{\alpha,0}}+ \sum_{j = 1}^l \lbk{\frac{i}{2} a_{-j\mathrm{B}} a_{j\mathrm{A}} + \frac{i}{2}a_{j\mathrm{B}} a_{-j\mathrm{A}}}, 
\eeq
which characterizes the pairing of the Majorana fermions between left-right symmetric sites such as $(j\B,-j\A)$ of the Kitaev chain. 

Before we discuss how to generate the entangled Majorana pairs, let us understand the physical meaning and eigenvalues of the conserved quantity, the $\mathcal{C}_\alpha$ operator. For this purpose, it is more intuitive to introduce
 the complex fermions using the Majoran fermions defined at the mirror symmetric sites:
\begin{eqnarray}
\alpha_0 &=& \frac{a_{0\B}+ia_{0\A}}{2}, \alpha_j = \frac{a_{j\B}+ia_{-j\A}}{2},\alpha_j^\prime = \frac{a_{-j\B}+ia_{j\A}}{2},\nonumber\\
\beta_0 &=& \frac{b_{0\A}+ib_{0\B}}{2}, \beta_j = \frac{b_{-j\A}+ib_{j\B}}{2}, \beta_j^\prime = \frac{b_{j\A}+ib_{-j\B}}{2},
\end{eqnarray}
where $\alpha_j$ and $\beta_j$ are from $(j\B,-j\A)$, $\alpha_j^\prime$ and $\beta_j^\prime$ are from $(j\A,-j\B)$. The non-zero anti-commutation relations of those complex fermion operators are $\{\alpha_j,\alpha_j^\dagger\} = \{\alpha^\prime_j,\alpha^{\prime\dagger}_j\}=\mathbbm{1}$, $\{\beta_j,\beta_j^\dagger\} = \{\beta^\prime_j,\beta^{\prime\dagger}_j\}=\mathbbm{1}$. 

The conserved quantity Eq.~\ref{eq:C_alpha} can be written as 
\beq 
\label{eq:C_number_representation}
\begin{aligned}
\mathcal{C}_\alpha &= \alpha_0^\dagger \alpha_0 - \frac12 + \sum_{j = 1}^{l} \lbk{\alpha^{\prime\dagger}_j \alpha^\prime_j+\alpha_j^\dagger \alpha_j -1}\\ 
&\equiv N_{\alpha,0}-\frac12 +\sum_{j = 1}^{l} \lbk{N_{\alpha,j}^\prime+ N_{\alpha,j}-1},\\
\end{aligned}
\eeq
where $N_{\alpha,j} = \alpha_j^\dagger \alpha_j$ and $N_{\alpha,j}^{\prime} = \alpha^{\prime\dagger}_j \alpha^\prime_j$, are the occupation numbers of the complex fermions $\alpha_j$ and $\alpha_j^\prime$. The complex fermions are made of the pairing of Majorana fermions between two distant sites. Note that the $b_{j\mu}$ operators and the corresponding complex fermion operators $\beta_j$ and $\beta_j^\prime$ are missing from the Hamiltonian Eq.~\ref{eq:hamil_kitaev_marjoana}. So $ib_{j\mu}b_{n\nu}$ are conserved quantities of the spin-1/2 Kitaev chain. One way to characterize these degrees of freedom is by identifying $N/2$ quantities $ib_{j\A}b_{j\B}$, each containing two levels. Hence there are $2^{N/2}$ degenerate states for the ground state of the Kitaev chain because of the $b_{i\mu}$ set of Majorana fermions. This is known as the Majorana zero modes in the Kitaev spin chain. The $b_{j\mu}$ Majorana fermion sets will get involved in the Hamiltonian as we move to the generalized XY chain in Sec.~\ref{sec:XY}, resulting in no degeneracy in the ground state of the generalized XY chain.

Following the complex fermion representation, Eq.~\ref{eq:C_number_representation}, the conserved quantity $\mathcal{C}_\alpha$ can take the eigenvalue from $-l-1/2$ to $l+1/2$. The maximum and minimum values correspond to the maximum pairing of Majorana fermions between left-right mirror symmetric sites. $\mathcal{C}_\alpha$ can be increased by consecutively pairing the two Majorana fermions on the 0-th unit cell, which increases the expectation value $\meanvl{\mathcal{C}_{\alpha,0}} = \meanvl{\frac{i}{2} a_{0\mathrm{B}} a_{0\mathrm{A}}}$, hence increasing $\meanvl{\mathcal{C}_\alpha}$. An ideal "pairing pulse" to generate the maximum bipartite entanglement 
is to apply the operator $(a_{0\B} - ia_{0\A})/2 = \alpha_0^\dagger$ on the central bond, which results in $\meanvl{\alpha_0^\dagger \alpha_0} \rightarrow 1$. The pairing pulse is represented as $\frac12 (\sigma_{0\B}^x \sigma_{0\A}^z + i \sigma_{0\A}^y)(\prod_{j<0}\sigma^z_{j\A}\sigma^z_{j\B})$ in the spin language, which involves sites on the left of $0\B$.

Here we demonstrate a process of generating long-range Majorana pairs. We start from the paramagnetic ordered state along the z-direction $|\psi_i\rangle = \prod_{j = 1}^N |\d\rangle$. The state has $\meanvl{\mathcal{C}_\alpha} =0$, which means it has no pairing of the Majorana fermions for the left and right symmetric sites. We then apply the pairing pulse $(a_{0\B} - ia_{0\A})/2$ on the state, gives $\frac12 (a_{0\B} - ia_{0\A})|\psi_i\rangle  = \frac12 (\sigma_{0\B}^x \sigma_{0\A}^z + i \sigma_{0\A}^y) (\prod_{j<0}\sigma^z_{j\A}\sigma^z_{j\B})|\d\d\rangle_{0\A,0\B}\otimes \prod_{j\neq 0}\lvert\d\d\rangle_{j\A,j\B} = \frac12 (-|\d\u\rangle+|\u\d\rangle)_{0\A,0\B}\otimes \prod_{j\neq 0}\lvert\d\d\rangle_{j\A,j\B}$. Note that the state $ \frac12 (a_{0\B} - ia_{0\A})|\psi_i\rangle$ has $\meanvl{\mathcal{C}_\alpha}  = 0.5$. Hence applying the pairing pulse $(a_{0\B} - ia_{0\A})/2$ changes $\meanvl{\mathcal{C}_{\alpha,0}}$ to $0.5$, making the Majorana fermions at the central sites pair together. For the next step we wait for the time evolution. The time evolution governed by the Kitaev Hamiltonian then creates an entangled state involving one more site each on the left and the right of the chain.


One can monitor the time evolution by monitoring $\meanvl{\mathcal{C}_{\alpha,0}} = -\frac12 \meanvl{\sigma^x_{0\A}\sigma^x_{0\B}}$. This value drops as the system evolves, while other contributions of $\meanvl{\mathcal{C}_\alpha}$ increase, indicating an entangled state between sites. We then monitor $\meanvl{\mathcal{C}_{\alpha,0}}$ until it drops by a tiny value, for example $0.001$. This choice is arbitrary and can take the range from $10^{-6}$ to $0.5$, we choose the value as $0.001$ because it applies to a wide range of parameters for the generalized XY chain, see sec.~\ref{sec:XY}. Furthermore, in real experiment one does not need to conduct measurement to know when to apply the pairing pulse. The state reaches rainbow state after $N/2$ pairing pulses are applied. The time interval between pulses should be finite, because the entanglement needs to be spread to further-distance sites related to the mirror-symmetry. For the case of $\delta \langle \mathcal{C}_{\alpha,0}\rangle \sim 10^{-6}$, the pulses are applied at $t = 0.06,0.23,0.65,1.3 \hbar/J$. We obtained numerically that the lower limit of time interval between the pulses is 0.04 $\hbar/J$. However, the precise value of the lower limit may depend on the size of the system, which is beyond the current study.

\begin{figure}
\includegraphics[width = 0.5 \textwidth]{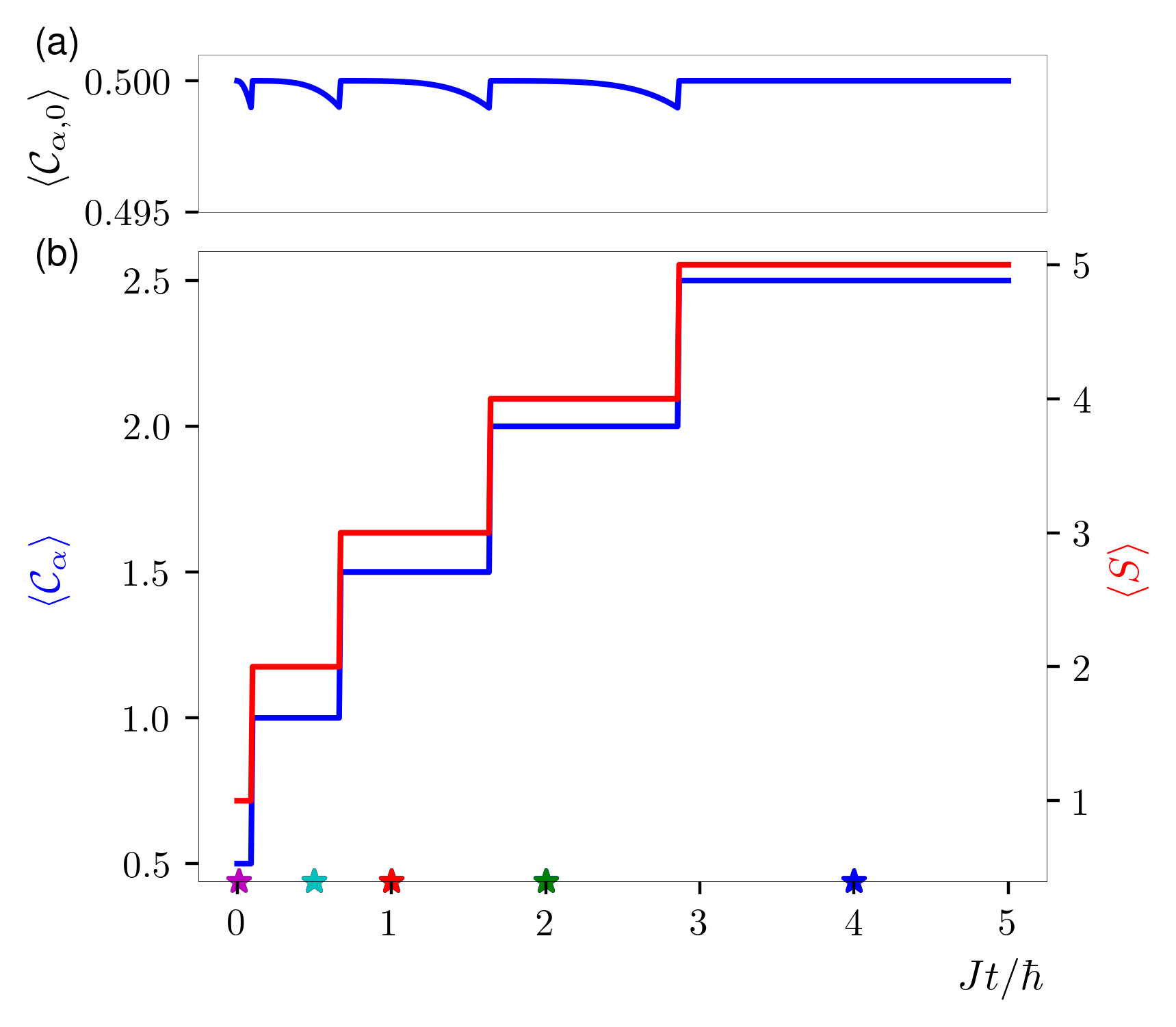}
\caption{
	Generating Majorana pairs by applying a pairing pulse $(a_{0\B} - ia_{0\A})/2$ on the central unit cell in the 10-site isotropic Kitaev chain.  (a) The time dependence of $\meanvl{\mathcal{C}_{\alpha,0}} = -\frac12 \meanvl{\sigma^x_{0\A} \sigma^x_{0\B} }$. Complex fermion pulse $\alpha_0^\dagger$ is applied at $Jt/\hbar = 0.09$, $0.66$, $1.63$, $2.86$, when $\meanvl{\mathcal{C}_{\alpha,0}}$ drops by $0.001$. The value goes back to $0.5$ after applying a pulse. (b) The time dependence of conserved quantities $\meanvl{\mathcal{C}_\alpha}$ and the bipartite entanglement entropy between the left and the right of the Kitaev chain $\meanvl{S}$ in the unit of $\log 2$. $\meanvl{\mathcal{C}_\alpha}$ increase by $0.5$ when the pulse is applied at the $0$-th unit cell, which is due to a new Majorana pair being created. The stars on the $t$-axis marks the time when the correlation function of the state is shown in Fig.~\ref{fig:kitaev_correlation}.
}
\label{fig:kitaev_quantities}
\end{figure}

In the example shown in Fig.~\ref{fig:kitaev_quantities}, we apply the pairing pulse $(a_{0\B} - ia_{0\A})/2$ one more time after $\meanvl{\mathcal{C}_{\alpha,0}}$ changes by the value of 0.001. The pairing pulse again brings the state of the central unit cell to $\frac12 (-|\d\u\rangle+|\u\d\rangle)_{0\A,0\B}$, with $\meanvl{\mathcal{C}_{\alpha,0}}  = 0.5$. And because the state is a many body state that mixes the sites, $\meanvl{\mathcal{C}_{\alpha} - \mathcal{C}_{\alpha,0}}$ also changes. In total $\meanvl{\mathcal{C}_{\alpha}}$ has a change of $0.5$. A new Majorana pair is generated between the site $1\A$ and $-1\B$. We then repeat the process of the time evolution and applying the pairing pulse until $\meanvl{\mathcal{C}_\alpha}$ reaches its maximum value of $l+1/2$. The final state contains maximum number of Majorana pairs between the left and right symmetric sites. The final state is protected by the mirror symmetry,
hence the further time evolution does not destroy the state.

To illustrate how the entangled pairs evolve in time,  exact diagonalization (ED) is performed for a 10-site isotropic Kitaev chain. Fig.~\ref{fig:kitaev_quantities}(a) shows the time dependence of $\meanvl{\mathcal{C}_{\alpha,0}} = -\frac12 \meanvl{\sigma^x_{0\A}\sigma^x_{0\B}}$. The pairing pulse $(a_{0\B} - ia_{0\A})/2$ is applied when $\meanvl{\mathcal{C}_{\alpha,0}}$ drops by $0.001$. Fig.~\ref{fig:kitaev_quantities}(b) shows the time dependence of conserved quantities $\meanvl{\mathcal{C}_\alpha}$ and the bipartite entanglement entropy  $S = -\Tr\rho \log \rho$, where $\rho$ is the reduced density matrix of the left or the right half of the chain. $\meanvl{\mathcal{C}_\alpha}$ increases by $0.5$ when the pairing pulse is applied, characterizing that one more Majorana pair is created in the chain. When $\meanvl{\mathcal{C}_\alpha}$ reaches the maximum, the left and the right sites of the chain are maximally paired to a direct product state of Majorana pairs, with the state $|\psi_f\rangle = \lbk{|\u\d\rangle - |\d\u\rangle}_{0\A,0\B} \otimes\prod_{j=1}^l \lbk{|\u\d\rangle - |\d\u\rangle}_{j\A,-j\B} \otimes \lbk{|\u\d\rangle - |\d\u\rangle}_{-j\A,j\B}$, which has $\frac{i}{2}\meanvl{a_{-j\B} a_{j\A}} = \frac{i}{2}\meanvl{a_{j\B} a_{-j\A}} = 1/2$. The entanglement entropy also has a leap when the pairing pulse is applied, since the pairing pulse creates a new entangled Majorana pair. When the final state is reached, the entanglement entropy between the left and the right also reaches its maximum, determined by the size of the system.

\begin{figure*}
\includegraphics[width = 1.0 \textwidth]{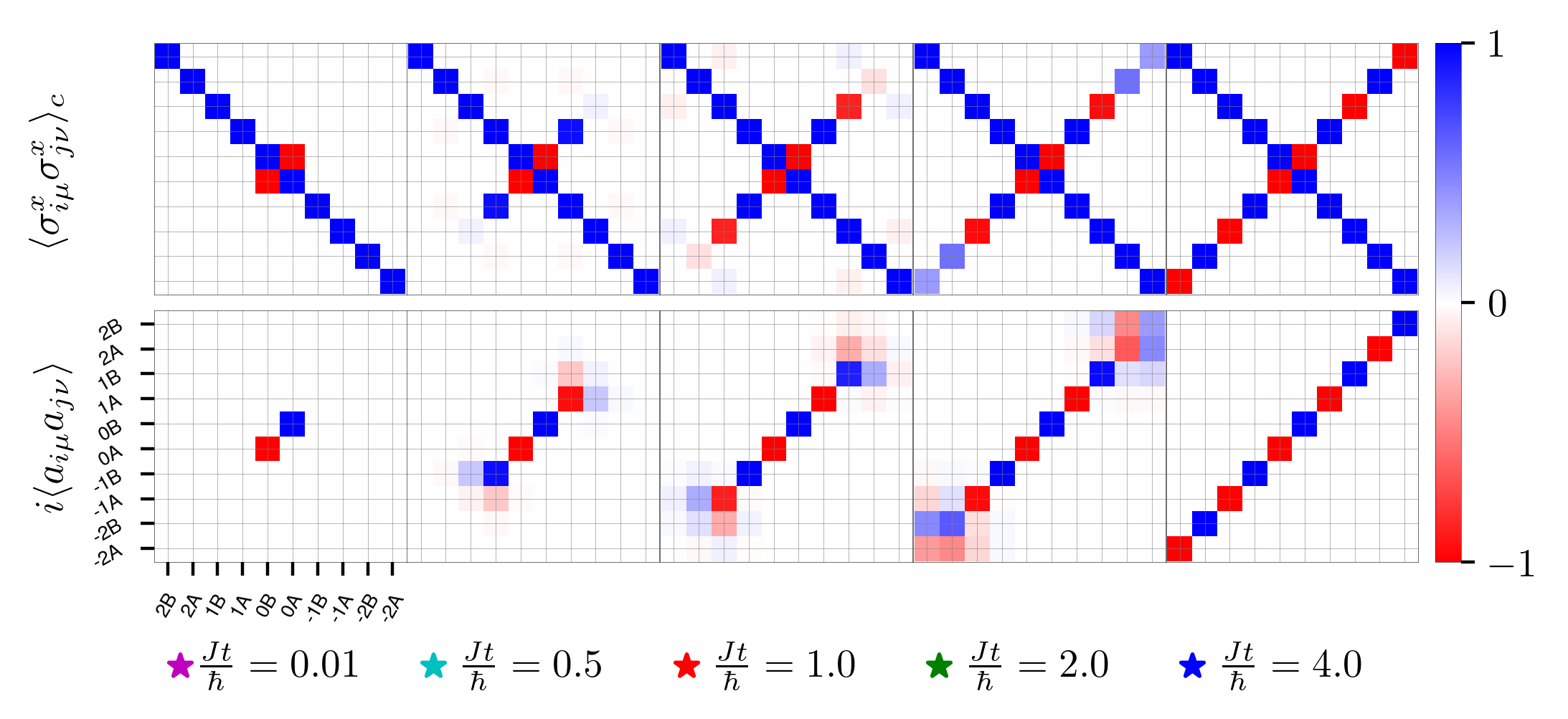}
\caption{
	The correlation functions of spins $\meanvl{\sigma^x_{i\mu} \sigma^x_{j\nu}}$ and the Majorana fermions $\meanvl{\frac{i}{2} a_{i\mu} a_{j\nu}}$ of states under the time evolution of the isotropic Kitaev Hamiltonian at $Jt/\hbar = 0.01$, $0.5$, $1.0$, $2.0$, $4.0$, also labelled as stars on $t$-axis in Fig.~\ref{fig:kitaev_quantities}. The time are chosen as a short time after applying each pulse.
}
\label{fig:kitaev_correlation}
\end{figure*}

We then inspect how the correlations of the state evolve in time. Fig.~\ref{fig:kitaev_correlation} shows the spin and Majorana correlations of the many body states at $Jt/\hbar = 0.01$, $0.5$, $1.0$, $2.0$, $4.0$, which are a short time after applying each pulse and are denoted by different colors of stars in the x-axis of Fig.~\ref{fig:kitaev_quantities}. The columns represent different time, and the rows respectively represent $\meanvl{\sigma_{i\mu}^x\sigma_{j\nu}^x}_c$ and $\meanvl{a_{i\mu}a_{j\nu}}$, where $\meanvl{\sigma_{i\mu}^x\sigma_{j\nu}^x}_c = \meanvl{\sigma_{i\mu}^x\sigma_{j\nu}^x} - \meanvl{\sigma_{i\mu}^x}\meanvl{\sigma_{j\nu}^x}$. The color on the diagonal entities show the correlation functions with the sites themselves. The anti-diagonal entities are between left-right symmetric sites of the chain. The $Jt/\hbar = 0.01$ state does not differ much from the correlation functions after applying the first pulse at $t=0$. After each pulse, we can see there is one more pair of Majorana fermions entangled, represented by the additional non-zero block on the anti-diagonal direction of the correlation matrix. After that the system evolves to mix one more site each on the left and the right, making the central unit cell entangled with the two further symmetric sites. Applying the pulse on the central unit cell again also changes the state of the entangled distant unit cells i.e., generating the Majorana pairs at these two cells. After a sequence of pulses, the state is fully entangled between the symmetric sites, as the $\meanvl{\sigma_{i\mu}^x\sigma_{j\nu}^x}$ and the $\meanvl{a_{j\mu}a_{n\nu}}$ matrix only has non-zero anti-diagonal elements.

The time taken to stand by for the next pulse to be applied takes the characteristic time of propagation. The time between the first pulse and second pulse approximately takes $\sim\hbar/J_{2y}$, the characteristic time for the state to evolve to a non-direct product state between the central unit cell and $(1\A,-1\B)$. In general, the time holding for evolution to entangle sites on the symmetric sites takes $\sim \hbar(N_y/J_{2y}+N_x/J_{1x})$, where $N_y$ and $N_x$ are the number of Y-bonds and X-bonds between one of the newly entangled sites and the central unit cell.

\section{Generalized XY chain~\label{sec:XY}}
\begin{figure}
	\includegraphics[width = 0.5\textwidth]{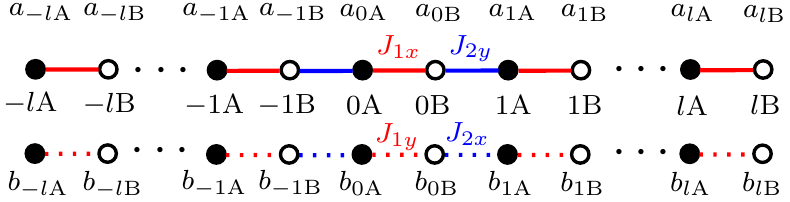}
	\caption{The description of the bond-dependent generalized XY chain in terms of the two independent Kitaev model, which can be represented as hopping of Majorana fermions $a_{i\mu}$'s and $b_{i\mu}$'s. Compared with the Kitaev chain (Fig.~\ref{fig:definition_kitaev}), another Kitaev chain made of $b_{i\mu}$ nearest neighbour hopping is added to the Hamiltonian, see Eq.~\ref{eq:hamil_XY}.}
	\label{fig:definition}
\end{figure}
In this section we apply our approach of generating Majorana pairs to the generalized XY spin chain. The generalized bond-dependent XY chain Hamiltonian can be written as 
\beq 
\label{eq:hamil_XY}
\begin{aligned}
\hamil =& \lbk{\frac{-J_{1x}}{4}} \sum_{j=-l}^l \sigma_{j\mathrm{A}}^x \sigma_{j\mathrm{B}}^x + \lbk{\frac{-J_{2y}}{4}} \sum_{j =-l}^{l-1} \sigma_{j\mathrm{B}}^y \sigma_{j+1, \mathrm{A}}^y \\
&+\lbk{\frac{-J_{1y}}{4}} \sum_{j=-l}^l \sigma_{j\mathrm{A}}^y \sigma_{j\mathrm{B}}^y + \lbk{\frac{-J_{2x}}{4}} \sum_{j =-l}^{l-1} \sigma_{j\mathrm{B}}^x \sigma_{j+1, \mathrm{A}}^x, \\
\end{aligned}
\eeq 
where the second line added to the Kitaev chain is the $J_{1y}$ and the $J_{2x}$ interactions. Note that for the Kitaev Hamiltonian, Eq.~\ref{eq:hamil_kitaev} one only needs one Majorana fermion $a_{i\mu}$ on each site. In fact, the hopping of the other sets of Majorana fermions, $b_{j\A}$ and $b_{j\B} $ contributes to the $(J_{1y},J_{2x})$ interaction. The generalized XY chain Hamiltonian, Eq.~\ref{eq:hamil_XY} can be written as 
\beq 
\label{eq:hamil_XY_marjoana}
\begin{aligned}
\hamil =& \frac{iJ_{1x}}{4} \sum_{j = -l}^{l} a_{j\mathrm{B}} a_{j\mathrm{A}}  + \frac{iJ_{2y}}{4} \sum_{j = -l}^{l-1}a_{j\mathrm{B}} a_{j+1,\mathrm{A}}  \\
&+\frac{iJ_{1y}}{4} \sum_{j = -l}b_{j\mathrm{A}}b_{j\mathrm{B}} + \frac{iJ_{2x}}{4} \sum_{j = -l}^{l-1}b_{j+1,\mathrm{A}}b_{j\mathrm{B}} .
\end{aligned}
\eeq 
This Majorana description of the generalized XY chain is plotted in Fig.~\ref{fig:definition}. Compared with the Hamiltonian of the Kitaev chain in Eq.~\ref{eq:hamil_kitaev} and Fig.~\ref{fig:definition_kitaev}, another Kitaev chain but with $b_{i\mu}$ hopping is added to the original Kitaev chain.

Similar to the $\mathcal{C}_\alpha$ conserved quantity defined in Eq.~\ref{eq:C_alpha}, there also exists a conserved quantity $[\mathcal{C}_\beta,\hamil]=0$ that connects $b_{i\mu}$ Majorana fermions between left and right,
\beq 
\label{eq:C_beta}
\begin{aligned}
    \mathcal{C}_\beta &= \frac{i}{2}b_{0\mathrm{A}} b_{0\mathrm{B}} + \sum_{j = 1}^{l} \lbk{\frac{i}{2} b_{j\mathrm{A}} b_{-j\mathrm{B}}+ \frac{i}{2}b_{-j\mathrm{A}}b_{j\mathrm{B}}}\\
    &= \beta_0^\dagger \beta_0 - \frac12 + \sum_{j =1}^l \lbk{\beta_j^{\prime\dagger} \beta_j^\prime + \beta_j^\dagger \beta_j } \\
    &\equiv N_{\beta,0}-\frac12 +\sum_{j = 1}^{l} \lbk{N_{\beta,j}^\prime+ N_{\beta,j}-1},\\
\end{aligned}
\eeq 
which measures the pairing of the $b_{j\mu}$ Majorana fermions between the left and the right. Note that $\{b_{n\mu}, a_{j\nu}\} =0$ for all $n,j,\mu,\nu$, so $b_{j\mu}$ and $a_{j\mu}$ are considered as two independent Majorana fermion sets in the generalized XY chain. As the generalized XY interaction Eq.~\ref{eq:hamil_XY_marjoana} does not have terms that mix $a_{j\mu}$ and $b_{j\nu}$ Majorana fermions, the two sets of the Majorana fermions behave independently. As a result, the Majorana pairs of $b_{i\mu}$ can also be generated by the same processes as in Sec.~\ref{sec:Kitaev} but with the pairing pulse switched to $\beta_0^\dagger =(b_{0\A}-ib_{0\B})/2$. Note that $\mathcal{C}_\beta$ varies synchronously as $\mathcal{C}_\alpha$ in our process because our initial state is a non-direct product state in the $\mathcal{C}_\alpha$ and $\mathcal{C}_\beta$ basis. It turns out that the Majorana pairing of the $a_{i\mu}$ state we generated in Sec.~\ref{sec:Kitaev} also corresponds to Majorana pairs of $b_{i\mu}$. We will discuss the relation of the $a$, $b$ Majorana pairs with the spin representation of Bell pairs in Sec.~\ref{sec:Bell}.

\begin{figure}
	\includegraphics[width = 0.499\textwidth]{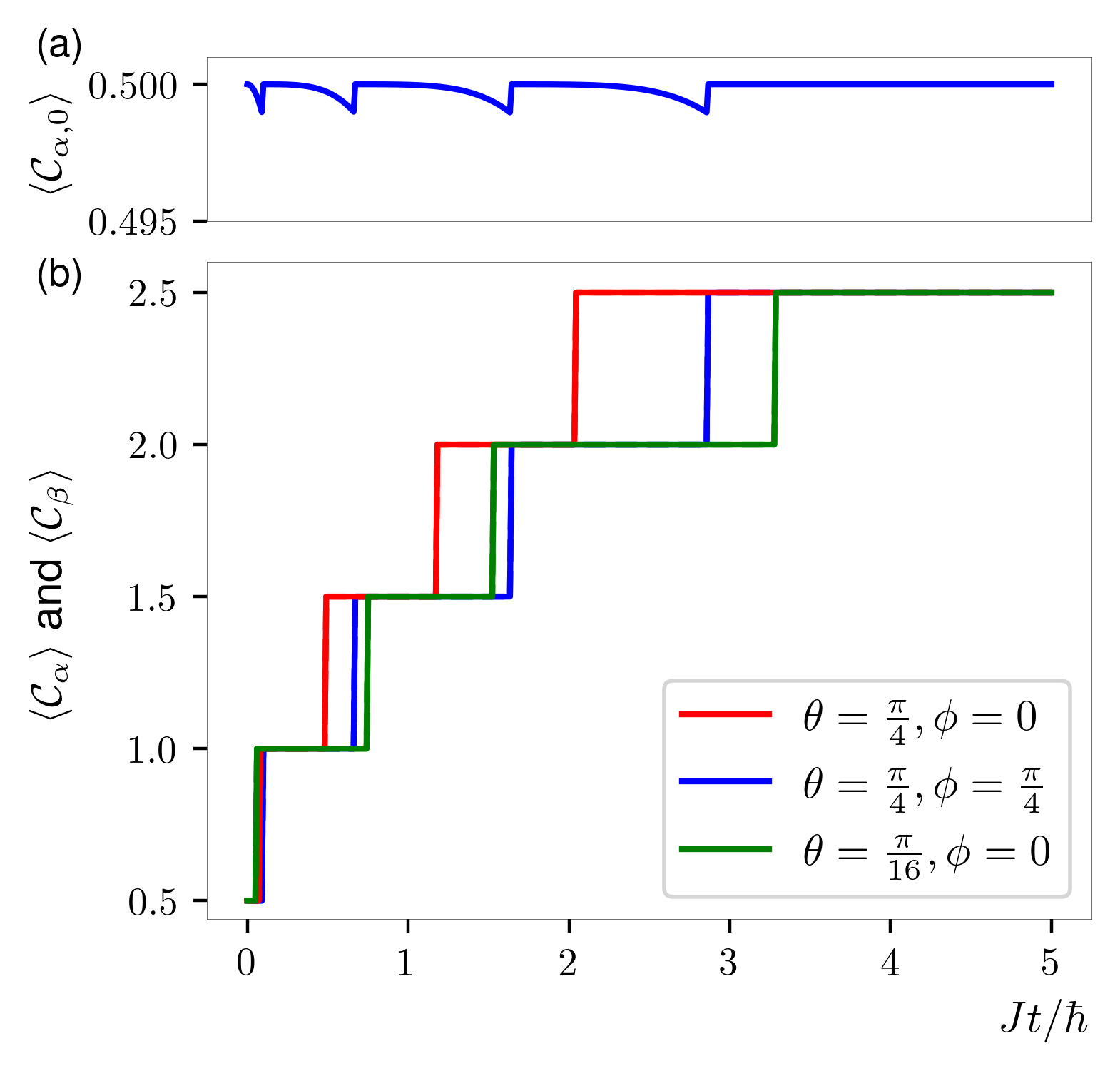}
	\caption{ Generating Majorana pairs by applying a pairing pulse for different parameter sets in 10-site generalized XY chains.  $\theta$ and $\phi$ determine the parameters in the Hamiltonian by $J_{1x} = J\sin\theta\cos\phi$, $J_{1y} = J\sin\theta\sin\phi$, $J_{2x} = J\cos\theta\sin\phi$ and $J_{2y} = J\cos\theta\cos\phi$. (a) The time dependence of $\meanvl{\mathcal{C}_{\alpha,0}} = -\frac12 \meanvl{\sigma^x_{0\A} \sigma^x_{0\B} }$ for $\theta = \pi/4,\phi = \pi/4$. Pairing pulse at the central unit cell is applied at $Jt/\hbar = 0.09$, $0.66$, $1.63$, $2.86$. (b) The time dependence of conserved quantities $\meanvl{\mathcal{C}_\alpha}$ and $\meanvl{\mathcal{C}_\beta}$ for different $(\theta,\phi)$. $\meanvl{\mathcal{C}_\alpha}$ and $\meanvl{\mathcal{C}_\beta}$ are equal and increase by $0.5$ when the pulse is applied at the $0$-th unit cell. 
	}
	\label{fig:quantities}
\end{figure}

We parameterize $J_{1x} = J\sin\theta\cos\phi$, $J_{1y} = J\sin\theta\sin\phi$, $J_{2x} = J\cos\theta\sin\phi$ and $J_{2y} = J\cos\theta\cos\phi$, where $0<\theta< \pi/2$ quantifies the ratio of the interaction between the 1-bond and the 2-bond, and $0\leq\phi\leq\pi/2$ defines the ratio between the two Kitaev chains. Our process of generating Majorana pairs does not work for $\theta = 0\text{ or }\pi/2$, as it corresponds to the two-site problem for each Kitaev chain, but our approach works for the pure Kitaev limit, $\phi = 0\text{ or }\pi/2$.
Using ED on a 10-site generalized XY chain, we show that the Majorana pairs are generated from an initial state of paramagnetic order for various ratios of $(J_{1x},J_{1y},J_{2x},J_{2y})$. 
Fig.~\ref{fig:quantities}(a) shows the time dependence of $\meanvl{\mathcal{C}_{\alpha,0}} = -\frac12 \meanvl{\sigma^x_{0\A}\sigma^x_{0\B}}$ with $(\theta,\phi) = (\pi/4,\pi/4)$ as an example.
The pairing pulse $(a_{0\B} - ia_{0\A})/2$ is applied when $\meanvl{\mathcal{C}_{\alpha,0}}$ drops by $0.001$. Fig.~\ref{fig:quantities}(b) shows the time dependence of conserved quantities $\meanvl{\mathcal{C}_\alpha}$ and $\meanvl{\mathcal{C}_\beta}$ in different parameter sets $(\theta,\phi)$. Finally when $\meanvl{\mathcal{C}_\alpha}$ and $\meanvl{\mathcal{C}_\beta}$ reach the maximum, long-range Majorana pairs are created. The final state is then 
$|\psi_f\rangle = \lbk{|\u\d\rangle - |\d\u\rangle}_{0\A,0\B} \otimes\prod_{i=1}^l \lbk{|\u\d\rangle - |\d\u\rangle}_{i\A,-i\B} \otimes \lbk{|\u\d\rangle - |\d\u\rangle}_{-i\A,i\B}$, which has the maximum value for both $\meanvl{\mathcal{C}_\alpha}$ and $\meanvl{\mathcal{C}_\beta}$.
It takes the total time of $\sim (N/2)\hbar(1/J_{1x}+1/J_{2y})$ to reach the maximally entangled rainbow-like state. For different parameters, $\phi$, the Kitaev limit $(\phi = 0,\pi/2)$ takes the minimum time.

\section{Kitaev to XY-chain and connection to entangled  Bell pairs~\label{sec:Bell}}
The conventional XY chain corresponds to $J_{1x}=J_{2x}=J_{1y}=J_{2y}$, which
 was studied by \citet{dutta2022generating}. It was shown that the entangled Bell pairs can be generated using a $\pi$-pulse.
On the other hand, in the Kitaev chain, we showed that the entangled Majorana pairs are created using the pairing pulse on the central bond.
As discussed above, the XY chain can be considered as two independent Kitaev chains made of two sets of Majorana fermions. 
In this section, we describe the relation between the Bell and Majorana pairs and show the time taken to reach the maximum entanglement for a
large parameter space of $\phi$ and $\theta$.

The entangled Majorana pairs are related to the Bell pairs, through the complex fermion number described by $\alpha_j^\dagger \alpha_j$ and $\beta_j^\dagger \beta_j$ that we introduced earlier. 
For each two sites there are two occupation number operators, hence there are $2^{(2\times N/2)} = 2^N$ states, which correspond to the $2^N$ spin states. 
An intuitive way to find the relation between the fermion occupation number and the spin states is to check a two-site problem. 

Consider only two sites $(0\A,0\B)$. The four eigenstates of $|N_{\alpha,0},N_{\beta,0}\rangle$ correspond to four entangled states in the spin, i.e., three triplets and one singlet, as shown in Fig.~\ref{five_states}. For the process of generating pairs, we start from a paramagnetic state
, which is a superposition of two complex fermion number eigenstates, $|\d\d\rangle = \frac12 (|0,1\rangle - |1,0\rangle)$ as shown in the center of Fig.~\ref{five_states}. 
Applying the pairing pulse, $\alpha_0^\dagger$ (the red solid line) generates a $|1,1\rangle$ state, which is a spin singlet Bell pair. 
Similarly applying a pairing pulse, $\beta_0^\dagger$ (the blue solid line) , also generates a $|1,1\rangle$ state. However, the Kitaev chain with only $J_{1x}$ and $J_{2y}$ does not involve $\beta$ fermions. The states that are not connected by the blue lines are responsible for the zero mode, and do not participate in the entangled Majorana pairs. On the other hand, in the conventional XY chain, either $\alpha^\dagger_0$ or $\beta^\dagger_0$  creates the entangled Bell pair.


\cpicn{1.0}{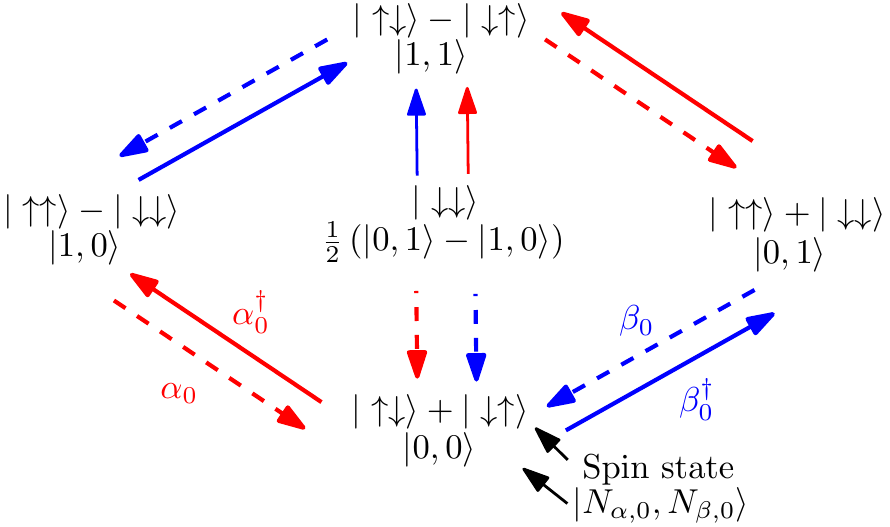}{The eigenstates of $N_{\alpha,0}$ and $N_{\beta,0}$ and their representation in the $\sigma^z$ spin basis for a $(0\A,0\B)$ two-site problem. The red and blue solid lines correspond to applying $\alpha_0^\dagger$ and $\beta_0^\dagger$, respectively, while the red and blue dashed lines correspond to applying $\alpha_0$ and $\beta_0$, respectively.}

\begin{figure}
	\centering
	\includegraphics[width = 0.5\textwidth]{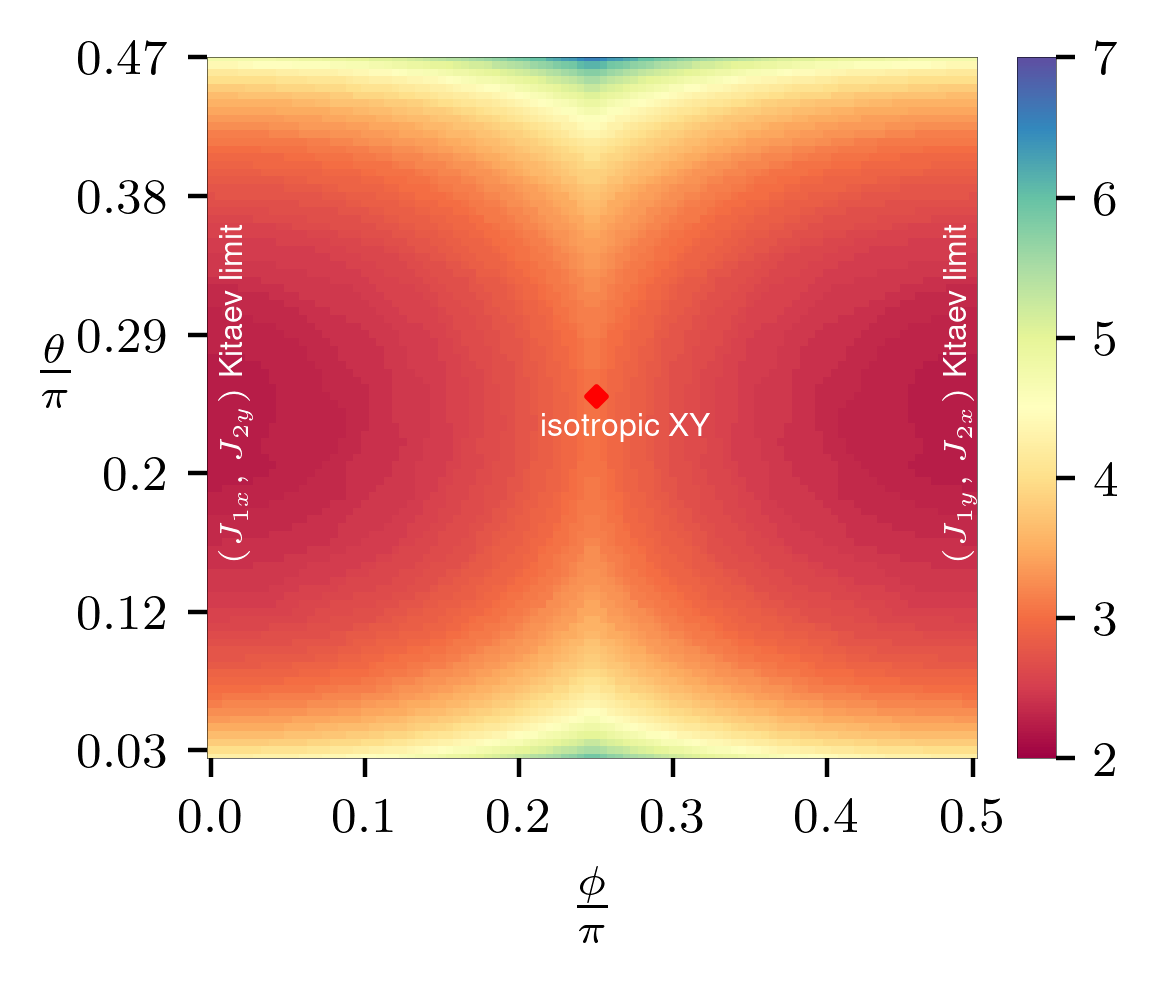}
	\caption{The time taken to reach the maximum sector of $\meanvl{\mathcal{C}_\alpha}$ for different $\theta$ and $\phi$, with the parameter range $(\theta = \lbk{\frac14 \pm 0.22}\pi$, $0\le \phi\le \frac{\pi}{2})$. }
	\label{fig:varying}
\end{figure}

For all the different ratios of $\{0<\theta<\pi/2,\ 0\le \phi\le \pi/2\}$, the state can reach a direct product state of Bell pairs with $\meanvl{\mathcal{C}_\alpha} = \meanvl{\mathcal{C}_\beta} = l+1/2$, with different time for different set of parameters. 
As the two Majorana fermions play equivalent roles in the system, one can come up with a similar process of generating Majorana pairs but with a $\beta_0^\dagger$ pairing pulse. As $\phi$ controls the magnitude between the two Kitaev models, $0<\phi<\pi/4$ has a larger value of the $a_{i\mu}$ hopping term $(J_{1x},J_{2y})$, while $\pi/4<\phi<\pi/2$ has a larger value of the $b_{i\mu}$ hopping term $(J_{2x},J_{1y})$. Since for $\phi/4<\phi<\pi/2$, it will take shorter time for the time evolution of the $b_{i\mu}$ hopping, it will take shorter time to reach the maximally entangled Bell pair state if one uses the pulse for $\phi/4<\phi<\pi/2$ instead. 

Fig.~\ref{fig:varying} shows the time taken to reach the maximally entangled Bell pairs for various $(\theta,\phi)$. The $\alpha_0^\dagger$ pairing pulse is applied to generate $a_{i\mu}$ Majorana pairs for $0<\phi\le\pi/4$, while the $\beta_0^\dagger$ pairing pulse is applied to generate $b_{i\mu}$ Majorana pairs for $\pi/4\le\phi\le\pi/2$. For a fixed $\theta$ which measures the ratio of the 1-bond and 2-bond, it takes the shortest time when $\phi$ reach $0$ or $\pi/2$, which is the corresponding Kitaev limit of the type of pulse applied. For a fixed $\phi$, it takes the shortest time when $\theta = \pi/4$, i.e., $J_{1\mu} = J_{2\mu}$ with $\mu=x,y$, which is the isotropic spin-1/2 Kitaev chain.

\section{Applying a spin-flip pulse to generate entanglement\label{sec:flip}}
In the previous sections we successfully generated the maximally entangled state using a pairing pulse on the central unit cell. The pairing pulse $\alpha_0^\dagger$ we applied is a linear combination of two spin-flip pulses. Thus a linear combination of $\alpha_0$ and $\alpha_0^\dagger$ can be a spin flip pulse. For example, the operator $\alpha_0+\alpha_0^\dagger = \sigma_{0\B}^x \sigma_{0\A}^z\prod_{j<0}\sigma_{j\A}^z\sigma_{j\B}^z$ can also increase the conserved quantity $\meanvl{\mathcal{C}_\alpha}$ when $\meanvl{\mathcal{C}_{\alpha,0}}<0$. This spin-flip operator corresponds to applying simultaneous $\pi$-pulses on half the qubits, which is feasible in cold atom experiments nowadays~\cite{dutta2022generating}. In fact, one can show that the pulse, $\alpha_0^\dagger+\alpha_0$, changes $\meanvl{\mathcal{C}_{\alpha,0}}$ to $-\meanvl{\mathcal{C}_{\alpha,0}}$, while it does not change other occupation numbers. As a result, one can wait for $\meanvl{\mathcal{C}_{\alpha,0}}$ to be negative and the spin-flip pulse can then increase the value of $\meanvl{\mathcal{C}_{\alpha}}$.
In this section we investigate the possible maximum $\meanvl{\mathcal{C}_\alpha}$ the system can reach by applying the spin-flip pulse. 



We start with a direct product of $\sigma^x_{j\mu}$ eigenstates, $|\psi_i\rangle = \prod_{j=1}^N (|\u\rangle +|\d\rangle)$. The occupation number form of $|\psi_i\rangle$ in the $(0\A,0\B)$ sub-Hilbert space is $|\psi_i\rangle \sim |0,0\rangle + |0,1\rangle$, which has $\meanvl{\mathcal{C}_\alpha} =-0.5$. 
Applying the spin-flip pulse, $(\alpha_0^\dagger + \alpha_0)$, sends the state to $|1,0\rangle+|1,1\rangle$, which has $\meanvl{\mathcal{C}_\alpha} =0.5$. To maximally increase $\meanvl{\mathcal{C}_\alpha}$, a pulse is applied when $\meanvl{\mathcal{C}_{\alpha,0}}$ reaches its negative minimum, i.e., $\meanvl{\mathcal{C}_{\alpha,0}}<0$ and $\partial_t \meanvl{\mathcal{C}_{\alpha,0}}=0$. After several pulses, $\meanvl{\mathcal{C}_{\alpha,0}}$ oscillates within a positive interval, hence applying the spin-flip pulse results in the decrease of $\meanvl{\mathcal{C}_{\alpha,0}}$. The system reaches the possibly maximally entangled state, when the spin-flip pulse cannot further increase $\meanvl{\mathcal{C}_{\alpha,0}}$.

\begin{figure}
    \includegraphics[width = 0.5\textwidth]{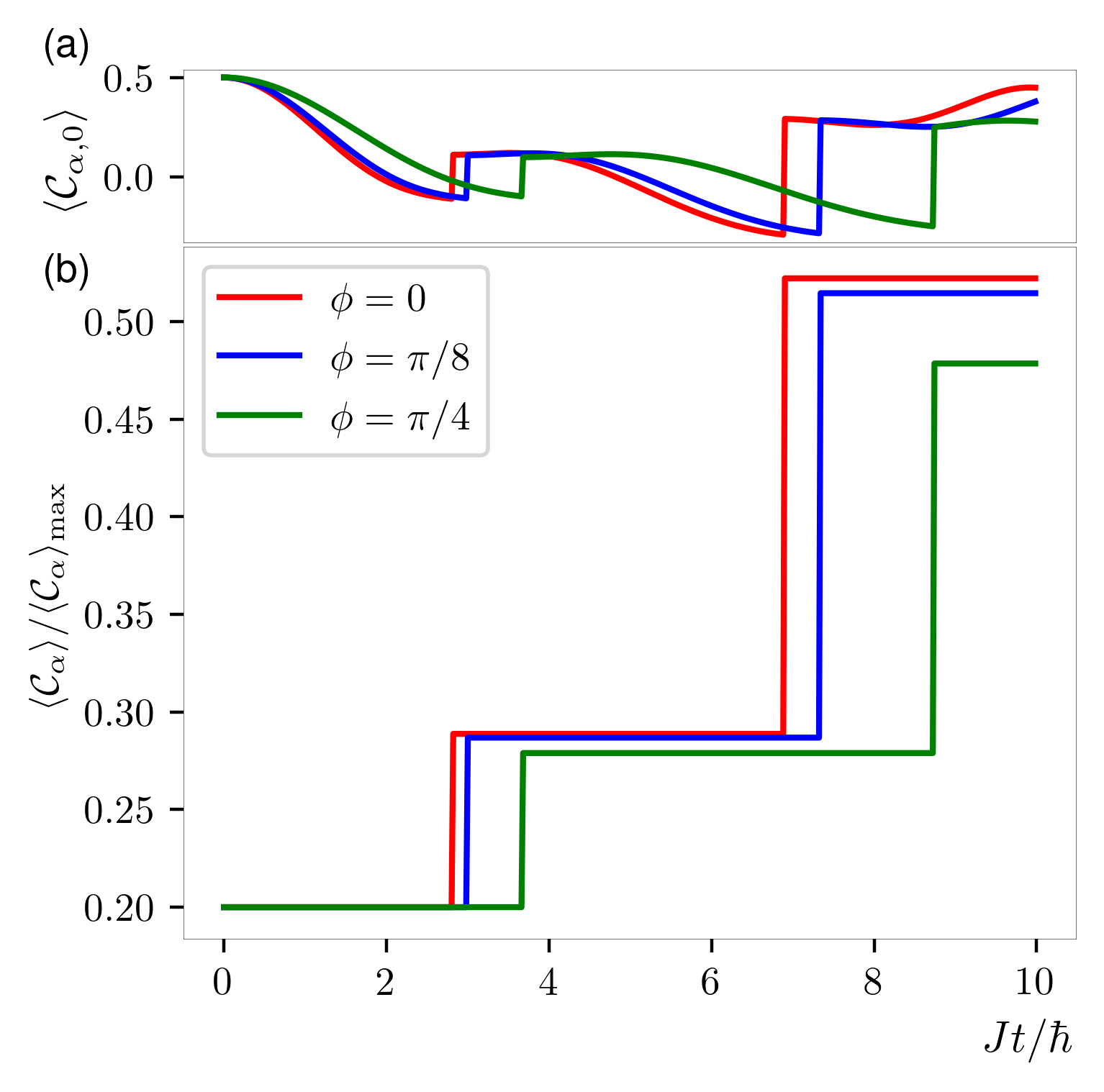}
    \caption{(a) The time evolution of $\meanvl{\mathcal{C}_{\alpha,0}}$ for various $\phi$ with $\theta = \pi/4$. The spin-flip pulse is applied when $\meanvl{\mathcal{C}_{\alpha,0}}$ reaches a negative minimum. (b) The time evolution of $\meanvl{\mathcal{C}_\alpha}/\meanvl{\mathcal{C}_\alpha}_{\mathrm{max}}$, $\sim 0.5$ of $\meanvl{\mathcal{C}_\alpha}_{\mathrm{max}}$ can be reached by applying the spin-flip pulse.
    }
    \label{fig:flipping_pulse}
\end{figure}

Numerical simulation is performed for a 10-site generalized XY chain, with $\theta = \pi/4$ for various $\phi$. Fig.~\ref{fig:flipping_pulse}(a) shows the evolution of $\meanvl{\mathcal{C}_{\alpha,0}}$. For different parameters one can reach a maximum of $\sim 0.5 (l+1/2)$. The $\phi=0$ line is the Kitaev limit, while $\phi = \pi/4$ is the isotropic XY chain. The spin-flip pulse does not yield the maximum entanglement in the XX chain, which is consistent with the results in Ref.~\cite{dutta2022generating}. Note that $\langle\mathcal{C}_\alpha\rangle$ measures the number of Majorana pairs, so part of the system can be described approximately by Majorana pairs.
The Kitaev limit gives the largest maximal $\meanvl{\mathcal{C}_\alpha}$, because our spin-flip pulse $\alpha_0^\dagger + \alpha_0$ promotes the $a_{j\mu}$ Majorana fermions pairing on the central unit cell, while for other parameter settings the Hamiltonian involves dynamics of the $b_{j\mu}$ Majorana fermions.

\section{Application to Twisted Kitaev chain \label{sec:twisted}}

\begin{figure}
    \centering
    \includegraphics[width = 0.3\textwidth]{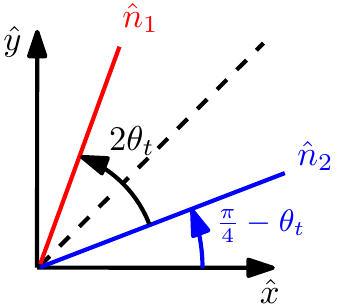}
    \caption{The interaction plane of the twisted Kitaev chain, $\hat{n}_1$ and $\hat{n}_2$ are two directions of the spin interaction on the 1 and 2 bond. When $\theta_t=0$ the model maps to Ising chain, and $\theta_t = \pi/4$ accounts for the Kitaev chain.}
    \label{fig:twisted_kitaev}
\end{figure}
 
\begin{figure}
    \centering
    \includegraphics[width = 0.5\textwidth]{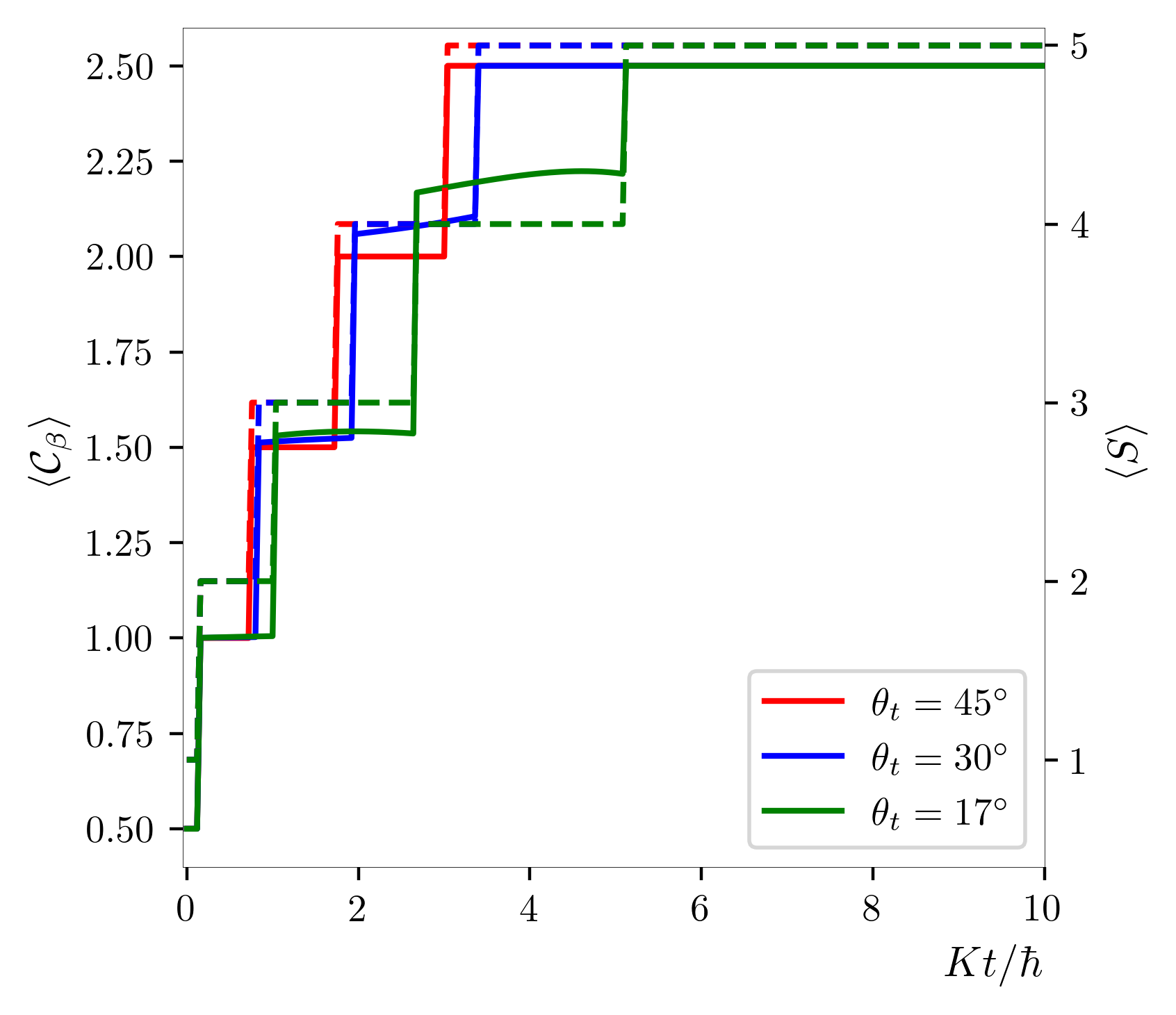}
    \caption{Creating Majorana pairs in the twisted Kitaev chain. The solid and dashed lines are the time evolution of $\meanvl{\mathcal{C}_\beta}$ and bipartite entanglement entropy $\meanvl{S}$ in the unit of $\log 2$ for various twisted angle $\theta_t$, respectively. } 
    \label{fig:twisted_quantities}
\end{figure}

We apply our theory to the recently proposed twisted Kitaev chain, CoNb$_2$O$_6$, described by the following Hamiltonian~\cite{morris2021duality,laurell2022spin},
\begin{equation}
\label{eq:twisted_hamil}
\hamil_t = -\frac{K}{4}\sum_{j} \lbk{\sigma_{j\A}^{\hat{n}_1} \sigma_{j\B}^{\hat{n}_1} + \sigma_{j\B}^{\hat{n}_2} \sigma_{j+1,\A}^{\hat{n}_2}},
\end{equation}
where $\sigma_{j\mu}^{\hat{n}} = \hat{n}\cdot \vec{\sigma}_{j\mu}$. $\hat{n}_1$ and $\hat{n}_2$ are the two directions of interaction for bond 1 and 2.  The two directions $\hat{n}_1$ and $\hat{n}_2$ are fully described by introducing the angle $2\theta_t$ between these two directions in the plane they determine. The relation between $\theta_t$ and the two directions is plotted in Fig.~\ref{fig:twisted_kitaev}. $\theta_t$ is experimentally measured as $17^\circ$ for the cobaltate chain, $\text{CoNb}_2\text{O}_6$.~\cite{morris2021duality} The Hamiltonian can be written as follows using the basis in Fig.~\ref{fig:twisted_kitaev},
\begin{equation}
\begin{aligned}
    &\hamil = -\frac{K}{4}\sum_j \left[\cos^2(\frac{\pi}{4}+\theta_t) \sigma^x_{j\A} \sigma^x_{j\B}+\sin^2(\frac{\pi}{4}+\theta_t) \sigma_{j\A}^y \sigma_{j\B}^y\right. \\
    &\left. + \frac{\cos 2\theta_t}{2}(\sigma_{j\A}^x\sigma_{j\B}^y+\sigma_{j\A}^y\sigma_{j\B}^x)\right]\\
    &+(-\frac{K}{4})\sum_j \left[\cos^2(\frac{\pi}{4}-\theta_t) \sigma^x_{j\B} \sigma^x_{j+1,\A}+\sin^2(\frac{\pi}{4}-\theta_t) \sigma_{j\B}^y \sigma_{j+1,\A}^y\right.\\
    &\left. + \frac{\cos 2\theta_t}{2}(\sigma_{j\B}^x\sigma_{j+1,\A}^y+\sigma_{j\B}^y\sigma_{j+1,\A}^x)\right].
\end{aligned}
\end{equation}
where the first and the third lines are the generalized XY model. The second and fourth lines are cross terms of $\sigma^x$ and $\sigma^y$, which accommodate non-Kitaev interactions that cannot be fully canceled out by a change of basis.  At $\theta_t = \pi/4$ the Hamiltonian goes to the Kitaev spin chain, with the cross terms becoming zero. Starting from the state $\prod_{j}|\d\d\rangle$, one can apply $\beta_0^\dagger$ when $\meanvl{\mathcal{C}_{\beta,0}}$ is less than $0.499$. This gives exactly the same result as the Kitaev chain in Sec.~\ref{sec:Kitaev}. With the cross terms turned on, although $\mathcal{C}_\alpha$ and $\mathcal{C}_\beta$ are not conserved under the whole Hamiltonian, we perform the same approach as for $\theta_t = \pi/4$. The results are shown in Fig.~\ref{fig:twisted_quantities}. For the non-Kitaev limit $\theta_t = 30^{\circ}$ and $\theta_t = 17^\circ$, after applying the pulse $\meanvl{\mathcal{C}_\beta}$ jumps to non-half-integer values. The time evolution afterward does not preserve $\meanvl{\mathcal{C}_\beta}$. But after $l$ pulses the state still reaches the rainbow-like state of Majorana pairs, which is the only state in the eigensector with the highest $\meanvl{\mathcal{C}_\beta}$ and $\meanvl{\mathcal{C}_\alpha}$ eigenvalue, even with the non-Kitaev interaction of the material comparable to the Kitaev interaction. The rainbow state is still protected by the mirror symmetry of the 1D chain in the twisted Kitaev chain, despite the fact that $\mathcal{C}_\beta$ is no longer conserved. The final state also has $\meanvl{\mathcal{C}_\alpha}=\meanvl{\mathcal{C}_\beta}=l+1/2$, and the wave function is the same as the state generated in the untwisted Kiteav chain.

\section{Effects of various perturbations on entanglement entropy\label{sec:stability}}
\begin{figure*}
    \centering
    \includegraphics[width = 1.0\textwidth]{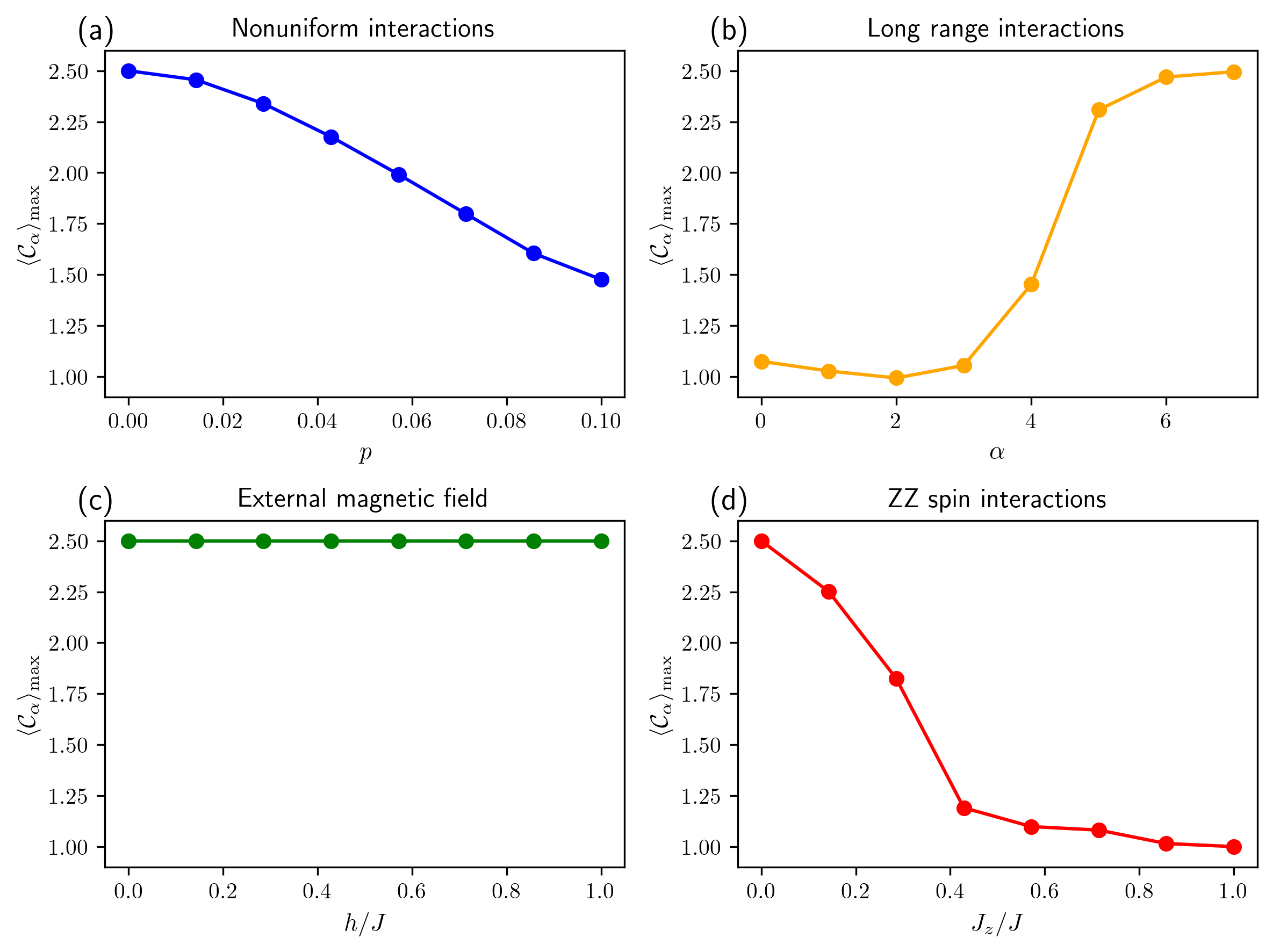}
    \caption{$\meanvl{\mathcal{C}_\alpha}_{\mathrm{max}}$ of 10-site Kitaev chain under (a)nonuniform bond interactions, (b)power law decay long-range interactions, (c)external magnetic field, (d)nearest neighbour ZZ interactions.When $\meanvl{\mathcal{C}_\alpha}_{\mathrm{max}}$ is 1, the corresponding entropy is $2 \log 2$, implying that two pairs are created.}
    \label{fig:perturbation}
\end{figure*}

As shown above, the twisted Kitaev chain contains not only the original Kitaev interaction (alternating $S^x S^x$ and $S^y S^y$) but also the cross terms such as $S^x S^y$. We showed that our protocol to create the rainbow state is valid despite the presence of other interactions. In this section, we consider other types of perturbations that may be present in solid-state materials. For example, a mirror-symmetry breaking perturbation like nonuniform interaction strength may affect the process of creating long-range entangled Majorana pairs.
Other perturbations, such as $(J_z/4)\sum_{j}^{N-1}\sigma_j^z\sigma_{j+1}^z$, i.e., $S^z S^z$ (ZZ) interactions, may also affect the structure of Majorana fermions. 

To investigate the effects of perturbations on the entanglement entropy, we consider four different perturbations which are added to the isotropic Kitaev chain $(J/4)\sum_j(\sigma_{j}^x \sigma_{j+1}^x+\sigma_{j}^y\sigma_{j+1}^y)$: (a) nonuniform bond interaction $\sum_j \frac{1}{4}J^p_{j,j+1} \sigma^x_j \sigma^x_{j+1}$, where $J^p_{j,j+1} = pj J$ with interaction strength varying by distance $j$ with a parameter $p$, (b) longer range interaction $\sum_j (J^L_{ij}/4)(\sigma^x_i\sigma^x_j+\sigma^y_i\sigma^y_j)$, where $J^L_{ij} = J/|i-j|^\gamma$, (c) external magnetic field $\sum_j h \sigma^z_j/2$, and (d) ZZ spin interaction $(J_z/4)\sum_{j}^{N-1}\sigma_j^z\sigma_{j+1}^z$. The results of maximum $\langle\mathcal{C}_\alpha \rangle $ are shown in Fig.~\ref{fig:perturbation}. 

As noted, a state with finite $\meanvl{\mathcal{C}_\alpha}$ contains entangled Majorana pairs. For nonuniform interactions, maximum $\meanvl{\mathcal{C}_\alpha}\sim 1.5$ for $p=0.1$, which means approximately 3 pairs are created in a 10-site chain for the largest $J^p_{j,j+1}\sim 0.9 J$. For long-range interactions, the limit when $\gamma\to\infty$ recap our previous results of creating $N/2$ Majorana pairs in the $N$-site system. On the other hand, the limit $\gamma=0$ means long-range interactions have the same magnitude of the nearest-neighbour spin interactions. In the $\gamma=0$ limit, we find $\meanvl{\mathcal{C}_\alpha}\sim 1.0$, indicating approximately 2 pairs are generated in the 10-site chain.
A homogeneous magnetic field along the $z$-direction, transpose to the Ising interaction direction, does not affect maximum entanglement at all. This is because the final state, i.e., the rainbow state is a simultaneous eigenstate of $\mathcal{H}_{\mathrm{total}} =( \text{Kitaev} + z\text{-direction magnetic field})$ and $\mathcal{C}_\alpha$ operator, even though these two operators do not commute. In other words, $\meanvl{\mathcal{C}_\alpha} = \meanvl{\text{rainbow state}|\mathcal{C}_\alpha|\text{rainbow state}}$ remain unchanged with time after the system reaches the rainbow state. In fact, the observation of the constant $\meanvl{\mathcal{C}_\alpha} = 2.5$ under the $z$-direction magnetic field confirms that we have reached the rainbow state independent of the field strength.

Finally, we find that under the ZZ spin interaction of up to $J_z/J = 1.0$, we still observe a finite $\meanvl{\mathcal{C}_\alpha}$, which approaches to $1$, implying two pairs similar to the longer range interaction $\gamma \rightarrow 0$ case. Overall, our protocol demonstrates robustness under a wide range of perturbations albeit the entanglement is smaller in some limits.

\section{Summary and Discussion \label{sec:Diss}}
To realize distributed quantum networks, creating long-range entangled pairs in many-body systems is desirable. However, entangled pairs of two distant qubits are difficult to generate due to interactions in many body systems. The rainbow state that maximally entangles the left and right part of a 1D chain was proposed.\cite{Ramirez_2014,vitagliano2010volume,PhysRevA.77.020303}
Recently a symmetry protected Bell pair entanglement such as the left-right mirror symmetry of the 1D XX spin chain using a $\pi$-pulse was suggested.\cite{dutta2022generating}

We investigated a way to generate long-range Majorana pairs in the spin-1/2 Kitaev chain, the 1D version of the honeycomb Kitaev model. 
We found that the maximally entangled rainbow-like Majorana pairing state can be reached, with a sequence of pairing pulses applied on the central unit cell of the system. 
The conserved quantity $\meanvl{\mathcal{C}_\alpha}$ associated with the mirror symmetry jumps by a half integer after a pulse, and reaches the maximum possible value in the Kitaev chain. 
We further generalized our method of creating long-range Majorana pairs into a generalized XY chain with bond-dependent interactions, which is described by two independent Kitaev chains. The results in the generalized XY chain are compared with that of the Kitaev chain. The Kitaev chain has the minimal time to reach the maximum entanglement entropy. This is because the process only involves one set of Majorana fermions, while the dynamics of the other set of Majorana fermions remains irrelevant to the process of generating Majorana pairs. 
%
We also studied the entangled states generated by a $\pi$-pulse. 
This spin-flip pulse would not generate the designated rainbow-like state, but can reach a fraction of $\meanvl{\mathcal{C}_\alpha}$ achieved by using the pair pulse and a significant bipartite entanglement entropy. 

Our theory can be applied to a twisted Kitaev chain, CoNb$_2$O$_6$, with an effective moment of $J_{\rm eff}=1/2$. This quasi-one dimensional chain has been known as one of the best examples of Ising ferromagnet. However recently it was suggested that that the bond-dependent Ising interaction, i.e., the Kitaev interaction with a tilted quantization axis, dubbed a twisted Kitaev chain model, describes its dynamics better.~\cite{morris2021duality} We applied our theory to the twisted Kitaev chain and found that the entangled Majorana pairs with finite entanglement entropy can also be generated by the pairing pulse. 
A challenge is applying the non-unitary pulse, which often requires post-selection process and takes considerably longer time than applying a unitary pulse. Thus how to implement the non-unitary pulse ~\citep{zheng2021universal,gingrich2004non} in real experiment needs to be explored in the future. For CoNb$_2$O$_6$, one can use the unitary pulse, which also results in the entangled state, even though it is not maximally entangled. The emergence of spin-spin correlations in this case under the spin-flip pulse signals the entangled Majorana pairs. 
Another direction that one could investigate in the future includes a higher-spin $S$ Kitaev ~\citep{Luo2021unusual,Sen2010PRB,jacob-spin-s,Stravropoulos2019micro} and generalized XY chains.









\section*{Acknowledgement}
We thank J. Gordon for insightful discussions. 
This work is supported by the Natural Sciences and Engineering Research Council of Canada (NSERC) and the Center for Quantum Materials at the University of Toronto. H.Y.K acknowledges the support by the Canadian Institute for Advanced Research (CIFAR) and the Canada Research Chairs Program. Computations were performed on the Niagara supercomputer at the SciNet HPC Consortium. SciNet is funded by: the Canada Foundation for Innovation under the auspices of Compute Canada; the Government of Ontario; Ontario Research Fund - Research Excellence; and the University of Toronto.

\bibliography{cites}


\appendix
\section{Dynamics of correlation functions \label{appendix:conserve}}
Here we list the results of the dynamics of the Majorana fermion correlation functions between the left-right symmetric sites in the Kitaev spin chain Fig.~\ref{fig:definition_kitaev}. It can be verified that 
\beq 
\label{eq:A1}
\begin{aligned}
&\partial_t \meanvl{\frac{i}{2}a_{-j\B} a_{j\A}} =\frac{i}{4\hbar} \meanvl{J_{1x} a_{-j\A}a_{j\A}+J_{2y} a_{-j+1,\A} a_{j\A} \\
& +J_{2y}a_{-j\B}a_{j-1,\B}+J_{1x} a_{-j\B}a_{j\B} }, \\
\end{aligned}
\eeq 
\beq
\label{eq:A2}
\begin{aligned}
&\partial_t\meanvl{\frac{i}{2} a_{-j\A}a_{j\B}} =\frac{i}{4\hbar} \meanvl{J_{2y}a_{j\B} a_{-j-1,\B} + J_{1x} a_{j\B}a_{-j\B}\\
&  +J_{1x}a_{-j\A}a_{j\A} + J_{2y} a_{j+1,\A}a_{j\B}}. 
\end{aligned}
\eeq
The evolution of the Majorana correlation function between $(0\A,0\B)$ follows
\beq 
\label{eq:zero_unit_cell}
\partial_t\meanvl{\frac{i}{2} a_{0\A}a_{0\B}} = \frac{i}{4\hbar} \meanvl{J_{2y}a_{0\B}a_{-1\B}+ J_{2y}a_{0\A} a_{1\A}},
\eeq 
the second derivative follows
\beq 
\label{eq:zero_unit_second}
\begin{aligned}
&\partial_t^2  \meanvl{\frac{i}{2} a_{0\A}a_{0\B}} =\frac{-iJ_{2y}}{8\hbar^2} \langle 2J_{2y} a_{0\B}a_{0\A} + 2J_{2y}a_{1\A}a_{-1\B} \\
& +J_{1x} a_{0\B} a_{-1\A}+ J_{1x} a_{0\A} a_{-1\B} + J_{1x}a_{1\A}a_{0\B} + J_{1x} a_{1\B}a_{0\A}\rangle .
\end{aligned}
\eeq
It turned out that the right hand side of Eq.~\ref{eq:A1}, Eq.~\ref{eq:A2} and Eq.~\ref{eq:zero_unit_cell} cancel out when the left hand side add up to $\mathcal{C}_\alpha$. Hence $\partial_t \meanvl{\mathcal{C}_\alpha} = \frac{i}{\hbar}\meanvl{[\hamil,\mathcal{C}_\alpha]} =0$. This commutation relation is independent of ratios between $J_{1x}$ and $J_{2y}$. The commutation relation is also independent of the sign of $J_{1x}$ and $J_{2x}$ i.e., ferro or anti-ferro interactions. The conservation of $\mathcal{C}_\alpha$ also allows different value for the $J_{1x}$ or $J_{2y}$ bond, but the left-right symmetry must exist. For example, the $J_{2y}$ between $0\B$ and $1\A$ sites can be different from the $J_{2y}$ between $1\B$ and $2\A$ sites, but must be the same as the $J_{2y}$ between $-1\B$ and $1\A$. Similarly, one can derive the conservation of $\mathcal{C}_\beta$. From Eq.~\ref{eq:zero_unit_second} one can verify the dynamic properties shown in Fig.~\ref{fig:kitaev_quantities} and Fig.~\ref{fig:quantities}.


\end{document}